\documentclass[12pt]{article}

\begin{document}
\begin{center}
 {\bf{\Large Black Holes in Models with Dilaton Field and Electric or Electric
 and Magnetic Charges}}
 \vspace{1cm}

E. Kyriakopoulos\footnote{E-mail: kyriakop@central.ntua.gr}\\
Department of Physics\\
National Technical University\\
157 80 Zografou, Athens, GREECE
\end{center}

\begin {abstract}
Exact static spherically symmetric charged black holes in four
dimensions are presented. One of them has only electric charge and
another electric and magnetic charges. In these solutions the
metric is asymptotically flat, has two horizons, irremovable
singularity only at $r=0$, and the dilaton field is singular only
at $r=0$. The solution with electric charge only is characterized
by three free parameters, the ADM mass, the electric charge and an
additional free parameter. It can be considered as a modification
of the GHS-GM solution obtained by changing the coupling between
dilaton and electromagnetic field. The general dyonic solution is
again characterized by three free parameters, the ADM mass, the
magnetic charge and an additional free parameter ,which is not the
electric charge. According to a definition of the no-hair
conjecture the solutions are "hairy".A very interesting special
case of the dyonic solution is characterized by three free
parameters, the ADM mass and the electric and the magnetic
charges. The solutions satisfy the dominant as well as the strong
energy condition outside and on the external horizon.

PACS number(s): 04.20.Jb, 04.70.Bw, 04.20.Dw

\end {abstract}

\section{Introduction}

In a previous work a black hole solution with dilaton and monopole
fields was found \cite{Ky1} and subsequently a generalization of
this solution was given \cite{Ky2}. It is therefore natural to try
to extend this work to the case where we have electric instead of
magnetic charge and to the case where we have both charges
electric and magnetic.

In this work we present two solutions one with electric charge
only and another with electric and magnetic charges, and we
consider special cases of these general solutions. The solutions
are exact and have some very nice features. All describe static
spherically symmetric black holes. Their metric is asymptotically
flat, has two horizons and irreducible singularity only at $r=0$.
The scalar field is singular only at $r=0$. Calculating the
eigenvalues of the energy-momentum tensor $T_{\mu\nu}$ of the
solutions we can show that all solutions satisfy the dominant as
well as the strong energy condition outside and on their external
horizon.

The solutions contain a number of parameters, which satisfy
certain relations. Therefore we can pick up a number of
independent parameters and express the rest in terms of them. The
independent parameters will tell us if the solution is "hairy" or
not. There are various formulations of the no-hair conjecture
\cite{Wh}-\cite{Nu}. According to one of them  \cite{Nu} "we say
that in a given theory there is black hole hair when the
space-time metric and the configurations of the other fields of a
stationary black hole solution are not completely specified by
conserved charges defined at asymptotic infinitely".

In Einstein-Maxwell theory the stationary asymptotically flat
solutions are determined by conserved charges defined at
asymptotic infinity and therefore they are not "hairy" \cite{He}.
The no-hair conjecture does not generalize to theories with
non-Abelian gauge fields coupled to gravity \cite{Vo}. Also
Einstein-Maxwell-Dilaton black hole solutions have nontrivial
dilaton field with an additional charge the dilaton charge. For
example the GHS-GM solution found by D. Garfinkle, G. T. Horowitz
and A. Strominger\cite{Ga} and previously by G. W. Gibbons and by
G. W. Gibbons and K. Maeda\cite{Gi} posses a dilaton charge, which
however can be expressed in terms of the mass and the magnetic
charge of the solution. In that sense the "hair" of the GHS-GM
solution is "secondary" \cite{Ka}.

The solution with electric charge only we have found has three
independent parameters and as such parameters we can chose the
Arnowitt-Deser-Misner (ADM) mass, the electric charge and an
additional free parameter. Therefore according to the above
definition it is a "hairy" solution. The "hair" is "primary".

The general dyonic solution and a special case of it have again
three free parameters, which can be the ADM mass the magnetic
charge and an additional free parameter. The electric charge is
determined from the magnetic charge and it is not a free
parameter. Again these solutions are "hairy" and their "hair" is
"primary". Another very interesting special case of the general
dyonic solution has as free parameters the ADM mass, the electric
charge and the magnetic charge, and its "hair" is "secondary".

The thermodynamic properties of the solutions will be studied
elsewhere. In the present work only their Hawking temperature is
given.

\section{Lagrangian and Equations of Motion}

Consider the action
\begin{equation}
\int d^{4}x \sqrt{-g} L = \int d^{4}x \sqrt{-g}
\{R-\frac{1}{2}\partial_{\mu}\psi\partial^{\mu}\psi-
f(\psi)F_{\mu\nu}F^{\mu\nu}\} \label{2-1}
\end{equation}
where $R$ is the Ricci scalar, $\psi$ is a dilaton field,
$f(\psi)$ is a function of $\psi$ and $F_{\mu\nu}$ is the
electromagnetic field. Since we shall consider only spherically
symmetric solutions having electric and magnetic charges or
electric charges only we shall write
\begin{equation}
F=\gamma(r) dr\wedge dt + Q_M sin\theta d\theta \wedge d\phi
\label{2-2}
\end{equation}
where $\gamma(r)$ is a function of $r$ only and $Q_M$ is the
magnetic charge of the solution. From this action we find the
following equations of motion for the electromagnetic field , the
dilaton field and the metric
\begin{equation}
(f F^{\mu\nu})_{;\mu}=0 \label{2-3}
\end{equation}

\begin{equation}
(\partial^{\rho}\psi)_{;\rho}-\frac{df}{d\psi}F_{\mu\nu}F^{\mu\nu}=0
\label{2-4}
\end{equation}

\begin{equation}R_{\mu\nu}=\frac{1}{2}\partial_{\mu}\psi\partial_{\nu}\psi+
2f(F_{\mu\sigma}{F_{\nu}}^{\sigma}-
\frac{1}{4}g_{\mu\nu}F_{\rho\sigma}F^{\rho\sigma})\label{2-5}
\end{equation}

We shall consider functions $f(\psi)$ such that if we introduce
them in Eqs (\ref{2-3})-(\ref{2-5}) we can solve these equations
and the solutions we get are asymptotically flat, have regular
horizons, have dilaton field singular only at $r=0$ and also have
electric and magnetic charges or electric charge only. We write
the metric in the form \cite{Ga}
\begin{equation}
ds^2=-\lambda^{2}dt^{2}+\lambda^{-2}dr^{2}+\xi^{2}d\Omega
\label{2-6}
\end{equation}
where $\lambda$ and $\xi$ are functions of $r$ only and
$d\Omega=d\theta^{2}+sin^{2}\theta {d{\phi}^{2}}$. From Eqs
(\ref{2-2}), (\ref{2-3}) and (\ref{2-6}) we get
\begin{equation}
(\xi^{2}f \gamma)'=0\label{2-7}
\end{equation}
where prime denotes differentiation with respect to $r$. From the
above relation we get
\begin{equation}
 \gamma=\frac{Q_E}{
 \xi^{2}f}  \label{2-8}
\end{equation}
where $Q_E$ is an integration constant. Also since $\psi=\psi(r)$
and $F_{\mu\nu}F^{\mu\nu}=-2\gamma^2+\frac{2Q_M^{2}}{\xi^{4}}$ the
dilaton Eq (\ref{2-4}), if we use Eq (\ref{2-8}), becomes

\begin{equation}
(\lambda^{2}\xi^{2}\psi')'=\frac{2}{\xi^2}\frac{d}{d\psi}(\frac{{Q_E}^{2}}{f}
+{Q_M}^2 f) \label{2-9}
\end{equation}

The non-vanishing components of the Ricci tensor of the metric
(\ref{2-6}) are $R_{00}$, $R_{11}$, $R_{22}$ and
$R_{33}=sin^{2}\theta{R_{22}}$, and for the first three components
we get respectively from Eqs (\ref{2-2}), (\ref{2-5}) (\ref{2-6})
and (\ref{2-8})

\begin{equation}
(\lambda^{2})''
+(\lambda^{2})'(\xi^{2})'\xi^{-2}=\frac{2}{\xi^4}(\frac{{Q_E}^{2}}{f}
+{Q_M}^2 f) \label{2-10}
\end{equation}
\[-(\lambda^{2})''\lambda^{-2}-2(\xi^{2})''\xi^{-2}-
(\lambda^{2})'(\xi^{2})'\lambda^{-2}\xi^{-2}+
[(\xi^{2})']^{2}\xi^{-4}=(\psi')^{2}\]
\begin{equation}-\frac{2}{\lambda^2 \xi^4}(\frac{{Q_E}^{2}}{f}
+{Q_M}^2 f) \label{2-11 }
\end{equation}
\begin{equation}
-[\lambda^{2}(\xi^{2})']'+2=\frac{2}{\xi^2}(\frac{{Q_E}^{2}}{f}
+{Q_M}^2 f) \label{2-12}
\end{equation}
Eqs  (\ref{2-9})-(\ref{2-12}) form a system of four equations for
the three unknowns $\lambda^{2}$, $\xi^{2}$ and $\psi$. Of course
to get a solution we have to specify first $f=f(\psi)$. This will
be done in such a way that  physically interesting solutions can
be found.

We have found previously \cite{Ky1} a solution with $\gamma=0$ and
$Q_M\neq0$ and later we generalized this solution \cite{Ky2}. In
the following sections we shall find a solution with $\gamma\neq0$
and $Q_M=0$ and another solution with $\gamma\neq0$ and
$Q_M\neq0$, and we shall consider special cases of these
solutions.

\section{Solution with Electric Field only}

Assume that

\begin{equation}
f(\psi)=\{g_{1}e^{(c+\sqrt{c^2+1})\psi}
+g_{2}e^{(c-\sqrt{c^2+1})\psi}\}^{-1} \label{3-1}
\end{equation}
where $c$, $g_1$ and $g_2$ are constants, and that

\begin{equation}
Q_M=0 \label{3-2}
\end{equation}

Then the system of Eqs (\ref{2-9})-(\ref{2-12}) has a solution
with

\[F=Q_E\{\frac{g_1}{r^2}e^{(c+\sqrt{c^2+1})\psi_0}+\frac{g_2}{(r+\alpha)^2}e^{(c
-\sqrt{c^2+1}\psi_0}\}dr \wedge dt\]
\begin{equation}
=\frac{1}{2Q_E}\{\frac{AB(\sqrt{c^2+1}-c)}{\sqrt{c^2+1}r^2}+
\frac{(\alpha-A)(\alpha-B)(\sqrt{c^2+1}+c)}{\sqrt{c^2+1}(r+\alpha)^2}\}dr
\wedge dt
 \label{3-3}
\end{equation}
\begin{equation}
\lambda^{2}=\frac{(r+A)(r+B)}{r(r+\alpha)}(\frac{r}{r+\alpha})^\frac{c}{\sqrt{c^2
+1}},\>\>\>\>\xi^{2}=
r(r+\alpha)(\frac{r+\alpha}{r})^\frac{c}{\sqrt{c^2+1}}\label{3-4}
\end{equation}
\begin{equation}
e^{\psi}=e^{\psi_{0}}(1+\frac{\alpha}{r})^\frac{1}{\sqrt{c^2+1}}
\label{3-5}
\end{equation}
where $A$, $B$, $\alpha$ and $\psi_{0}$ are integration constants,
provided that the following relations are satisfied
\begin{equation}
g_{1}=\frac{AB(\sqrt{c^2+1}-c)}{2Q_E^{2}\sqrt{c^2+1}}e^{-(c+\sqrt{c^2+1})\psi_{0}}
\label{3-6}
\end{equation}
\begin{equation}
 g_{2}=\frac{(\alpha-A)(\alpha-B)(\sqrt{c^2+1}
+c)}{2Q_E^{2}\sqrt{c^2+1}}e^{-(c-\sqrt{c^2+1})\psi_{0}}
\label{3-7}
\end{equation}
From Eqs (\ref{2-6}) and (\ref{3-4}) we get
\[ds^{2}=-\frac{(r+A)(r+B)}{r(r+\alpha)}(\frac{r}{r+\alpha})^\frac{c}{\sqrt{c^2
+1}}dt^{2}+\frac{r(r+\alpha)}{(r+A)(r+B)}(\frac{r+\alpha}{r})^\frac{c}{\sqrt{c^2
+1}}dr^{2}\]
\begin{equation}
+r(r+\alpha)(\frac{r+\alpha}{r})^\frac{c}{\sqrt{c^2+1}}d{\Omega}
\label{3-8}
\end{equation}
Our solution is given by Eqs (\ref{3-3}) and
(\ref{3-5})-(\ref{3-8}).

Eq (\ref{3-3}) indicates that we have a point charge located at
the origin and a charge distribution, which is spherically
symmetric with respect to the origin.

Eqs (\ref{3-5}) and (\ref{3-8}) have appeared in a previous work
\cite{Ky2} and therefore some of the arguments presented there
hold also in the present case. From Eq (\ref{3-8}) we get
asymptotically
\begin{equation}
-g_{00}=1-\frac{\alpha(1+\frac{c}{\sqrt{c^2+1}})-A-B}{r} +
O(r^{-2})\label{3-9}
\end{equation}
which means that the solution is asymptotically flat and its ADM
mass $M$ is given by
\begin{equation}
2M=\alpha(1+\frac{c}{\sqrt{c^2+1}})-A-B \label{3-10}
\end{equation}
Eq (\ref{3-5}) tells us that $\psi_{0}$ is the asymptotic value of
$\psi$. Also we shall make the choice
\begin{equation}
\alpha>0,\>\>\>\>\> A<0,\>\>\>\>\> B<0 ,\label{3-11}
\end{equation}
for which $\psi$ is singular only at $r=0$ and for which we have
\begin{equation}
g_{1}>0,\>\>\>\>\> g_{2}>0,\>\>\>\>\> M>0\label{3-12}
\end{equation}

The solution has the integration constants $\alpha$, $A$, $B$,
$\psi_{0}$ and $Q_E$, which for given $g_{1}$, $g_{2}$ and $c$
must satisfy Eqs (\ref{3-6}) and (\ref{3-7}). Therefore only three
of them are independent. Introducing the ADM mass $M$ by the
relation  (\ref{3-10}) we can take $M$, $Q_E$ and $\psi_{0}$ as
independent parameters. Therefore our solution has arbitrary mass,
arbitrary electric charge and an additional arbitrary parameter.
Thus it is a "hairy" solution according to the definition given in
Ref. \cite{Nu}. The "hair" is "primary".

If (\ref{3-11}) hold our metric is singular at $r=-A$ and  $r=-B$
but not at $r=-\alpha$. However the Ricci scalar $R$ and the
curvature scalar $R_{\mu\nu\rho\sigma}R^{\mu\nu\rho\sigma}$ are
not singular at $r=-A$ and  $r=-B$. Indeed we have  \cite{Ky2}
\begin{equation}
R=\frac{\alpha^{2}(r+A)(r+B)}{2(c^2+1)r^{3}(r
+\alpha)^{3}}(\frac{r}{r+\alpha})^\frac{c}{\sqrt{c^2+1}}\label{3-13}
\end{equation}
\begin{equation}
R_{\mu\nu\rho\sigma}R^{\mu\nu\rho\sigma}=
\frac{P(r,\alpha,A,B)}{4(c^2+1)^2
r^{6}(r+\alpha)^{6}}(\frac{r}{r+\alpha})^\frac{2c}{\sqrt{c^2+1}}
\label{3-14}
\end{equation}
where $P(r,\alpha,A,B)$ is a complicated polynomial of $r$,
$\alpha$ $A$ and $B$. Therefore  we have  an irremovable
singularity only at $r=0$.

Proceeding as in Ref. \cite{Ky1} and \cite{Ky2} we can introduce
Eddington- Finkelstein type coordinates and we can show that we
have a black hole solution with two horizons located at $r=-A$ and
$r=-B$.

The energy-momentum tensor $T_{\mu\nu}$ of our solution is given
by
\[T_{\mu\nu}=\partial_{\mu}\psi\partial_{\nu}\psi
+4f F_{\mu\rho}{F_{\nu}}^{\rho} -g_{\mu
\nu}\{\frac{1}{2}\partial_{\rho}\psi\partial^{\rho}\psi +f F_{\rho
\sigma}F^{\rho \sigma}\}\]
\[=\frac{\alpha^{2}}{(c^2+1)r^{2}(r+\alpha)^{2}}(\delta_{\mu
r}\delta_{\nu
r}-\frac{1}{2}g_{\mu\nu}g^{rr})+\frac{2}{r^2(r+\alpha)^2}\{\frac{AB(\sqrt{c^2+1}-
c)}{\sqrt{c^2+1}}
\]
\[\times(\frac{r+\alpha}{r})^{\frac{c+\sqrt{c^2+1}}{\sqrt{c^2+1}}}
+\frac{(\alpha-A)(\alpha-B)(\sqrt{c^2+1}+c)}{\sqrt{c^2+1}}(\frac{r+
\alpha}{r})^{\frac{c-\sqrt{c^2+1}}{\sqrt{c^2+1}}}\}\]
\[
\times\{-\frac{r(r+\alpha)}{(r+A)(r+B)}(\frac{r+
\alpha}{r})^{\frac{c}{\sqrt{c^2+1}}}\delta_{\mu r}\delta_{\nu r}\]
\begin{equation}
+\frac{(r+A)(r+B)}{r(r+\alpha)}(\frac{r}{r+
\alpha})^{\frac{c}{\sqrt{c^2+1}}}\delta_{\mu t}\delta_{\nu
t}+\frac{g_{\mu\nu}}{2} \} \label{3-15}
\end{equation}
Calculating the eigenvalues of $T_{\mu\nu}$ we can show that it
satisfies the dominant as well as the strong energy condition
outside and on the external horizon.

The Hawking temperature $T_H$ of our solution if $A<B<0$ and
$\alpha>0$ is given by \cite{Ho}
\begin{equation}
T_H=\frac{|{\lambda^2}'(-A)|}{4\pi} = \frac{(B-A)(-A)^{-1
+\frac{c}{\sqrt{c^2+1}}}(\alpha-A)^{-1-\frac{c}{\sqrt{c^2+1}}}}{4\pi}
\label{3-16}
\end{equation}

We argued before that we can take $M$, $Q_E$ and $\psi_{0}$ as
independent parameters of the solution. In this case using Eqs
(\ref{3-6}), (\ref{3-7}) and (\ref{3-10}) we can express $A$, $B$
and $\alpha$ in terms of $M$, $Q_E$ and $\psi_0$. For $A\leq B<0$,
$\alpha>0$ and assuming for simplicity that $c=0$ we get
\[A=\frac{1}{2M}\{Q^2_E(g_2 e^{-\psi_0}-g_1 e^{\psi_0})-2M^2\]
\begin{equation}
-\sqrt{[Q^2_E(g_2 e^{-\psi_0}-g_1 e^{\psi_0})-2M^2]^2-8M^2Q^2_Eg_1
e^\psi_0} \} \label{3-17}
\end{equation}
\[B=\frac{1}{2M}\{Q^2_E(g_2 e^{-\psi_0}-g_1 e^{\psi_0})-2M^2\]
\begin{equation}
+\sqrt{[Q^2_E(g_2 e^{-\psi_0}-g_1 e^{\psi_0})-2M^2]^2-8M^2Q^2_Eg_1
e^\psi_0} \}  \label{3-18}
\end{equation}
\begin{equation}
\alpha=\frac{Q^2_E}{M}(g_2 e^{-\psi_0}-g_1 e^{\psi_0})\label{3-19}
\end{equation}
where
\begin{equation}
 2M^2
\geq
Q^2_E(\sqrt{g_2}e^{-\frac{\psi_0}{2}}+\sqrt{g_1}e^{\frac{\psi_0}{2}})^2\label{3-20}
\end{equation}
\begin{equation}
g_2 e^{-\psi_0}-g_1 e^{\psi_0}>0 \label{3-20'}
\end{equation}

We have for the extremal solution $A=B<0$ and $\alpha>0$. This
happens if Eq (\ref{3-20}) with the equality sign and Eq.
(\ref{3-20'}) hold. In this case we get
\begin{equation}
A=B=-\sqrt{2g_1}Q_E e^\frac{\psi_0}{2} \label{3-21}
\end{equation}
Also if $A>0$, $B>0$ and $\alpha>0$ the metric has a naked
singularity at $r=0$. This happens if
\begin{equation}
2M^2<
Q^2_E(\sqrt{g_2}e^{-\frac{\psi_0}{2}}-\sqrt{g_1}e^{\frac{\psi_0}{2}})^2
 \>\>\>\> \mbox{and}\>\>\>\>g_2
e^{-\psi_0}-g_1 e^{\psi_0}>0 \label{3-22}
\end{equation}
Finally if $g_1=1$ and $g_2=0$ our solution reduces to the GHS-GM
solution for electric charge found by D. Garfinkle, G. T. Horowitz
and A. Strominger\cite{Ga}, and by M. Rakhmanov\cite{Ra}, and
previously by G. W. Gibbons and by G. W. Gibbons and K.
Maeda\cite{Gi}.

\section{Dyonic Solution}

Let us take
\[f_\pm(\psi)=g'_{1}e^{(c+\sqrt{c^2+1})\psi}
+g'_{2}e^{(c-\sqrt{c^2+1})\psi}\]
\begin{equation}
\pm\sqrt{[g'_{1}e^{(c+\sqrt{c^2+1})\psi}
+g'_{2}e^{(c-\sqrt{c^2+1})\psi}]^2+g_3}
 \label{4-1}
\end{equation}
where $c$, $g'_1$, $g'_2$ and $g_3$ are real constants. Then the
system of Eqs (\ref{2-9})-(\ref{2-12}) has a solution with
\[F_\pm=\frac{Q_E}{\xi^2
f_\pm}dr\wedge dt+Q_M sin\theta d\theta\wedge
d\phi=\frac{Q_M^2f_\mp}{Q_E \xi^2}dr\wedge dt+Q_M sin\theta
d\theta\wedge d\phi\]
\[=\frac{Q_M^2
e^{c\psi_0}}{Q_E}\{\frac{g'_{1}e^{\sqrt{c^2+1}\psi_0}}{r^2}+
\frac{g'_{2}e^{-\sqrt{c^2+1}\psi_0}}{(r+\alpha)^2}
\mp[\frac{g'^2_{1}e^{2\sqrt{c^2+1}\psi_0}}{r^4}+
\frac{g'^2_{2}e^{-2\sqrt{c^2+1}\psi_0}}{(r+\alpha)^4}\]
 \[+\frac{2g'_1 g'_2}{r^2(r+\alpha)^2}+\frac{g_3 e^{-2c\psi_0}}{r^2(r+
 \alpha)^2}(\frac{r}{r+\alpha})^{\frac{2c}{\sqrt{c^2+1}}}
]^{\frac{1}{2}} \}dr\wedge dt+Q_Msin\theta d\theta\wedge dt \]
\[=\frac{1}{4\sqrt{c^2+1}Q_E}\{\frac{AB(\sqrt{c^2+1}-c)}{r^2}+
\frac{(\alpha-A)(\alpha-B)(\sqrt{c^2+1}+c)}{(r+\alpha)^2} \]
\[\mp[\frac{A^2B^2(\sqrt{c^2+1}-c)^2}{r^4}+
\frac{(\alpha-A)^2(\alpha-B)^2(\sqrt{c^2+1}+c)^2}{(r+\alpha)^4}+\]
\begin{equation}
\frac{2AB(\alpha-A)(\alpha-B)}{r^2(r+\alpha)^2} -\frac{16Q^2_E
Q^2_M
(c^2+1)}{r^2(r+\alpha)^2}(\frac{r}{r+\alpha})^\frac{2c}{\sqrt{c^2+
1}}]^\frac{1}{2}\}dr\wedge dt+Q_Msin\theta d\theta\wedge dt
 \label{4-2}
\end{equation}
and with $\lambda^2$, $\xi^2$ and $\psi$ given by Eqs (\ref{3-4})
and \ref{3-5}), where $A$, $B$, $\alpha$, $\psi_0$, $Q_E$ and
$Q_M$ are constants, which are connected with the given constants
$g'_1$, $g'_2$, $g_3$ and $c$ with the relations
\begin{equation}
g'_{1}=\frac{AB(\sqrt{c^2+1}-c)}{4Q_M^{2}\sqrt{c^2+1}}e^{-(c+\sqrt{c^2+1})\psi_{0}}
\label{4-3}
\end{equation}
\begin{equation}
 g'_{2}=\frac{(\alpha-A)(\alpha-B)(\sqrt{c^2+1}
+c)}{4Q_M^{2}\sqrt{c^2+1}}e^{-(c-\sqrt{c^2+1})\psi_{0}}
\label{4-4}
\end{equation}
\begin{equation}
g_3=-\frac{Q^2_E}{Q^2_M} \label{4-5}
\end{equation}
The charges $Q_E$ and $Q_M$ are connected by the previous
relation, which means that it is not possible to take arbitrary
values both. Our most general dyonic solution is given by Eqs
(\ref{3-5}), (\ref{3-8}) and (\ref{4-2})-(\ref{4-5})

The ADM mass $M$ of the solution is given by Eq (\ref{3-10}) and
for the choice  (\ref{3-11}), which we shall make again, $\psi$ is
singular only at $r=0$ and we get
\begin{equation}
g'_{1}>0,\>\>\>\>\> g'_{2}>0,\>\>\>\>\> M>0\label{4-6}
\end{equation}

The solution has a number of constants namely the magnetic charge
$Q_M$ and the integration constants $A$, $B$, $\alpha$, $\psi_0$
and $Q_E$, which for given $g'_1$, $g'_2$, $g_3$ and $c$ must
satisfy Eqs (\ref{4-3})-(\ref{4-5}). Therefore only three of them
are independent and if we take into account Eq. (\ref{3-10}) we
can choose $M$, $Q_M$ and $\psi_0$ as independent parameters. Thus
the solution has arbitrary mass, arbitrary magnetic charge and an
additional arbitrary parameter. Therefore it is a "hairy" solution
according to the definition given in Ref. \cite{Nu}. The "hair" is
"primary".

Since the Ricci scalar $R$ and the curvature scalar
$R_{\mu\nu\rho\sigma}R^{\mu\nu\rho\sigma}$ of the solution are
given by Eqs (\ref{3-13}) and (\ref{3-14}) the singularities at
$r=-A$ and $r=-B$  are coordinate singularities and only the
singularity at $r=0$ is irremovable. We have a black hole solution
with two horizons at $r=-A$ and $r=-B$.

The energy-momentum tensor $T_{\mu\nu}$ of our solution is given
by
\[T_{\mu\nu}=\partial_{\mu}\psi\partial_{\nu}\psi
+4f_{\pm} F_{\mu\rho}{F_{\nu}}^{\rho} -g_{\mu
\nu}\{\frac{1}{2}\partial_{\rho}\psi\partial^{\rho}\psi +f_{\pm}
F_{\rho \sigma}F^{\rho \sigma}\}\]
\[=\frac{\alpha^{2}}{(c^2+1)r^{2}(r+\alpha)^{2}}\delta_{\mu
r}\delta_{\nu r} +\frac{h_\mp} {\sqrt{c^2+1}r(r+\alpha)}\]
\[\times\{\frac{(r+A)(r+B)}{r(r+\alpha)}(\frac{r}{r+
\alpha})^{\frac{c}{\sqrt{c^2+1}}}\delta_{\mu t}\delta_{\nu
t}-\frac{r(r+\alpha)}{(r+A)(r+B)}(\frac{r+
\alpha}{r})^{\frac{c}{\sqrt{c^2+1}}}\delta_{\mu r}\delta_{\nu r}
\}\]
\[+\frac{h_\pm}{\sqrt{c^2+1}}(\frac{r}{r+\alpha})^\frac{c}{\sqrt{c^2+1}}
(\delta_{\mu\theta}\delta_{\nu\theta}+sin^2\theta\delta_{\mu\phi}\delta_{\nu\phi})
-\frac{g_{\mu\nu}}{2}\{\frac{\alpha^2(r+A)(r+B)}{(c^2+1)r^3(r+\alpha)^3}\]
\begin{equation}
-\frac{h_\mp}{\sqrt{c^2+1}r(r+\alpha)}+\frac{h_\pm}{\sqrt{c^2+1}r(r+
\alpha)}(\frac{r}{r+\alpha})^{\frac{2c}{\sqrt{c^2+1}}}
 \} \label{4-7}
\end{equation}
where
\[h_\pm=4\sqrt{c^2+1}Q_M^2(\frac{r+\alpha}{r})^{\frac{c}{\sqrt{c^2
+1}}}\{g'_1\frac{e^{(c+\sqrt{c^2+1})\psi_0}}{r^2}
+g'_2\frac{e^{(c-\sqrt{c^2+1})\psi_0}}{(r+\alpha)^2}\]
\begin{equation}
\pm[\{g'_1\frac{e^{(c+\sqrt{c^2+1})\psi_0}}{r^2}
+g'_2\frac{e^{(c-\sqrt{c^2+1})\psi_0}}{(r+\alpha)^2}\}^2+g_3(\frac{r}{r+\alpha})^{\frac{2c}{\sqrt{c^2+1}}}
]^{\frac{1}{2}} \}
 \label{4-8}
\end{equation}
or
\[h_\pm=\frac{AB(\sqrt{c^2+1}-c)}{r^2}(\frac{r+\alpha}{r})^{\frac{c}{\sqrt{c^2+1}}}
+\frac{(\alpha-A)(\alpha-B)(\sqrt{c^2+1}+c)}{(r+\alpha)^2}\]
\[\times(\frac{r}{r+
\alpha})^{\frac{c}{\sqrt{c^2+1}}}\pm\{[\frac{AB(\sqrt{c^2+1}-c)}{r^2}
(\frac{r+ \alpha}{r})^{\frac{c}{\sqrt{c^2+1}}}\]
\begin{equation}
+\frac{(\alpha-A)(\alpha-B)(\sqrt{c^2+1}+c)}{(r+\alpha)^2}(\frac{r}{r+
\alpha})^{\frac{c}{\sqrt{c^2+1}}} ]^2
-\frac{16(c^2+1)Q_E^2Q_M^2}{r^2(r+\alpha)^2} \}^{\frac{1}{2}}
\label{4-9}
\end{equation}
Calculating the eigenvalues of $T_{\mu\nu}$ we can show that it
satisfies the dominant as well as the strong energy condition
outside and on the external horizon.

The Hawking temperature $T_H$ of our solution if $A<B<0$ and
$\alpha>0$ is given by Eq. (\ref{3-16}).

If we chose $M$, $Q_M$ and $\psi_0$ as free parameters using Eqs
(\ref{3-10}) (\ref{4-3})and (\ref{4-4}) we can express $A$, $B$
and $\alpha$ in terms of $M$, $Q_M$ and $\psi_0$. To simplify the
calculations we shall assume that $c=0$ and we shall call $g''_1$
and $g''_2$ the expressions $g'_1$ and $g'_2$ of Eqs
(\ref{4-3})and (\ref{4-4}) for $c=0$ i. e.
\begin{equation}
g'_1(c=0)\equiv g''_1=\frac{AB}{4Q_M^2}e^{-\psi_0}\>\>\>\>\>\>\>\>
g'_2(c=0) \equiv
g''_2=\frac{(\alpha-A)(\alpha-B)}{4Q_M^2}e^{\psi_0} \label{4-10}
\end{equation}
Then for $A\leq B<0$ and $\alpha>0$ we find that
\[A=\frac{1}{M}\{Q^2_M(g''_2 e^{-\psi_0}-g''_1 e^{\psi_0})-M^2\]
\begin{equation}
-\sqrt{[Q^2_M(g''_2 e^{-\psi_0}-g''_1
e^{\psi_0})-M^2]^2-4M^2Q^2_Mg''_1 e^{\psi_0}} \} \label{4-11}
\end{equation}
\[B=\frac{1}{M}\{Q^2_M(g''_2 e^{-\psi_0}-g''_1 e^{\psi_0})-M^2\]
\begin{equation}
+\sqrt{[Q^2_M(g''_2 e^{-\psi_0}-g''_1
e^{\psi_0})-M^2]^2-4M^2Q^2_Mg''_1 e^{\psi_0}} \}  \label{4-12}
\end{equation}
\begin{equation}
\alpha=\frac{2Q^2_M}{M}(g''_2 e^{-\psi_0}-g''_1
e^{\psi_0})\label{4-13}
\end{equation}
and that we must have
\begin{equation}
M^2 \geq
Q^2_M(\sqrt{g''_2}e^{-\frac{\psi_0}{2}}+\sqrt{g''_1}e^{\frac{\psi_0}{2}})^2
\label{4-14'}
\end{equation}
\begin{equation}
g''_2 e^{-\psi_0}-g''_1 e^{\psi_0}>0\label{4-15'}
\end{equation}

We get the extremal solution if $A=B<0$ and $\alpha>0$. This
happens if relation (\ref{4-14'}) with the equality sign and
relation (\ref{4-15'}) hold. Also if $A>0$, $B>0$ and $\alpha>0$
the metric has a naked singularity at $r=0$. This happens if
\begin{equation}
M^2 <
Q^2_M(\sqrt{g''_2}e^{-\frac{\psi_0}{2}}-\sqrt{g''_1}e^{\frac{\psi_0}{2}})^2
\label{4-14}
\end{equation}

A very interesting dyonic solution is obtained if in the above
general dyonic solution given by Eqs (\ref{4-1}), (\ref{3-5}),
(\ref{3-8}) and (\ref{4-2})-(\ref{4-5}) we put
\begin{equation}
c=0,\>\>\>\>\>\>g''_2=\frac{1}{2}\>\>\>\>\>\>\mbox{and}\>\>\>\>\>\>g_3=-4g''_1
g''_2 \label{4-15}
\end{equation}
where $g''_1$ and $g''_2$ denote the values of $g'_1$ and $g'_2$
if $c=0$. Then Eq (\ref{4-1}) gives
\begin{equation}
f_{-}(\psi)=e^{-\psi} \label{4-16}
\end{equation}
and Eqs (\ref{3-5}), (\ref{3-8}) and (\ref{4-2})-(\ref{4-5})
become
\begin{equation}
e^{\psi}=e^{\psi_{0}}(1+\frac{\alpha}{r})
 \label{4-17}
\end{equation}

\begin{equation}
ds^{2}=-\frac{(r+A)(r+B)}{r(r+\alpha)}dt^{2}
+\frac{r(r+\alpha)}{(r+A)(r+B)}dr^{2} +r(r+\alpha)d{\Omega}
 \label{4-18}
\end{equation}

\begin{equation}
F_{-}=\frac{Q_Ee^{\psi_0}}{r^2}dr \wedge dt +Q_M sin\theta d\theta
\wedge d\phi
 \label{4-19}
\end{equation}

\begin{equation}
\frac{AB
e^{-\psi_0}}{2{Q_E}^2}=\frac{(\alpha-A)(\alpha-B)e^{\psi_0}}{2{Q_M}^2}=1
 \label{4-20}
\end{equation}

That is if in the Lagrangian of Eq. (\ref{2-1}) we have
$f(\psi)=f_{-}(\psi)=e^{-\psi} $ Eqs (\ref{4-17})-(\ref{4-19}),
where the constants $A$, $B$, $\alpha$, $\psi_0$, $Q_E$ and $Q_M$
satisfy the relations (\ref{4-20}), give a dyonic solution. Eq.
(\ref{4-19}) tells us that the electromagnetic field is created by
a point electric charge $Q_E e^{\psi_0}$ and a point magnetic
charge $Q_M$, which are located at the origin.

The ADM mass of the solution is given by Eq. (\ref{3-10}) with
$c=0$ i.e.
\begin{equation}
2M=\alpha-A-B \label{4-22}
\end{equation}
Also $\psi_0$ is the asymptotic value of $\psi$ and for the choice
\begin{equation}
\alpha>0\mbox{,}\>\>\>\>\>\>A<0\mbox{,}\>\>\>\>\>\>B<0
\label{4-23}
\end{equation}
which is consistent with the constraint Eqs (\ref{4-20}) and for
which $2M>0$, the metric is singular at $r=0$, $r=-A$ and $r=-B$.
However the Ricci scalar $R$ and the curvature scalar
$R_{\mu\nu\rho\sigma}R^{\mu\nu\rho\sigma}$, which are given by Eqs
(\ref{3-13}) and (\ref{3-14}) with $c=0$, are singular only at
$r=0$. This means that the singularities at $r=-A$ and $r=-B$ are
coordinate singularities and only at $r=0$ we have an irremovable
singularity. The solution is a black hole solution with two
horizons located at $r=-A$ and $r=-B$. Also since $\alpha>0$ the
dilaton field $\psi$ is singular only at $r=0$.

The action (\ref{2-1}) if $f(\psi)$ is the $f_{-}(\psi)$ of Eq.
(\ref{4-16}) is invariant under the transformation
\begin{equation}
g_{\mu\nu}\rightarrow
g_{\mu\nu}\mbox{,}\>\>\>\>\>\>\psi\rightarrow\psi
+b\mbox{,}\>\>\>\>\>\>F_{\mu\nu}\rightarrow
e^{\frac{b}{2}}F{\mu\nu} \label{4-24}
\end{equation}
where b is a constant.This invariance leads to the conserved
dilaton current
\begin{equation}
J_\mu=\partial_\mu \psi+2 e^{-\psi}F_{\mu\nu}A^\nu
 \label{4-25}
\end{equation}
where $A^\nu$ is the vector potential, from which we get the
conserved dilaton charge
\begin{equation}
D=\frac{1}{4\pi}\int_S J_\mu dS^\mu=-\alpha
 \label{4-26}
\end{equation}
where the integral is over a two-sphere at spatial infinity. In
deriving the above expression for the dilaton charge we have
assumed that $A^t$ vanishes at infinity. Since the transformation
(\ref{4-24}) changes $\psi_0$ we can give it any value we want,
for example the value zero. Therefore $\psi_0$ cannot be
considered as a parameter of the model. This means that we have
only three free parameters, since we have the parameters $A$, $B$,
$\alpha$, $Q_M$ and $Q_E$ which must satisfy Eqs (\ref{4-20}). We
shall choose as free parameters the mass $M$ and the electric and
magnetic charges $Q_E$ and $Q_M$.

If we define $Q'_E$ and $Q'_M$ by the relations
\begin{equation}
Q'_E=Q_E e^\frac{\psi_0}{2}\mbox{,}\>\>\>\>Q'_M=Q_M
e^\frac{-\psi_0}{2}
 \label{4-27}
\end{equation}
and solve Eqs (\ref{4-20}) and (\ref{4-22}) for $A$, $B$ and
$\alpha$ in terms of $M$, $Q'_E$ and $Q'_M$ assuming that $A \leq
B<0$ and $\alpha>0$ we find
\begin{equation}
A=\frac{1}{2M}\{{Q'_M}^2-{Q'_E}^2-2M^2-\sqrt{({Q'_M}^2-{Q'_E}^2-2M^2)^2-8M^2{Q'_E}^2}
\}
 \label{4-28}
\end{equation}
\begin{equation}
B=\frac{1}{2M}\{{Q'_M}^2-{Q'_E}^2-2M^2+\sqrt{({Q'_M}^2-{Q'_E}^2-2M^2)^2-8M^2{Q'_E}^2}
\}
 \label{4-29}
\end{equation}
\begin{equation}
\alpha=\frac{1}{M}({Q'_M}^2-{Q'_E}^2)
 \label{4-30}
\end{equation}
and that we must have
\begin{equation}
\sqrt{2}M \geq \mid Q'_ M \mid +\mid Q'_ E \mid \label{4-31}
\end{equation}
and
\begin{equation}
\mid Q'_ M \mid -\mid Q'_ E \mid >0 \label{4-32}
\end{equation}
Using Eqs (\ref{4-26})and (\ref{4-30}) we can express the dilaton
charge $D$ in terms of $Q'_E$ and $Q'_M$. We get
\begin{equation}
D=\frac{1}{M}({Q'_E}^2-{Q'_M}^2)
 \label{4-33}
\end{equation}

We have for the extremal solution $A=B<0$ and $\alpha>0$. This
happens if Eq. (\ref{4-31}) with the equality sign and Eq.
(\ref{4-32}) hold. Also if $A>0$, $B>0$ and $\alpha>0$ the metric
has a naked singularity at $r=0$. This happens if
\begin{equation}
\sqrt{2}M < \mid Q'_ M \mid -\mid Q'_ E \mid\label{4-34}
\end{equation}

If $Q'_M=0$ the previous solution becomes the the GHS-GM solution
for electric charge only. Also if $Q'_E=0$ this solution becomes
\begin{equation}
ds^2=(1-\frac{2M}{r+\frac{{Q'_M}^2}{r}})dt^2+(1-\frac{2M}{r+
\frac{{Q'_M}^2}{r}})^{-1}dr^2+r(r+\frac{{Q'_M}^2}{r})d\Omega
\label{4-35}
\end{equation}
\begin{equation}
e^\psi=e^{\psi_0}\frac{r+\frac{{Q'_M}^2}{r}}{r} \label{4-36}
\end{equation}
and if we replace $r$ by $r'=r+\frac{{Q'_M}^2}{r}$ we get the
GHS-GM solution for magnetic charge.

The energy-momentum tensor $T_{\mu\nu}$ of the solution given by
Eqs  (\ref{4-17})-(\ref{4-20}) is
\[T_{\mu\nu}=\partial_{\mu}\psi\partial_{\nu}\psi
+4e^{-\psi} F_{\mu\rho}{F_{\nu}}^{\rho} -g_{\mu
\nu}\{\frac{1}{2}\partial_{\rho}\psi\partial^{\rho}\psi +e^{-\psi}
F_{\rho \sigma}F^{\rho \sigma}\}\]
\[=\frac{\alpha^2}{r^2(r+\alpha)^2}\delta_{\mu r}\delta_{\nu r}
+\frac{4{Q'_E}^2}{r(r+\alpha)^3}\{
\frac{(r+A)(r+B)}{r(r+\alpha)}\delta_{\mu t}\delta_{\nu t}
-\frac{r(r+\alpha)}{(r+A)(r+B)}\delta_{\mu r}\delta_{\nu r} \}\]
\begin{equation}
+\frac{4{Q'_M}^2}{r^2}(\delta_{\mu\theta}\delta_{\nu\theta}
+sin^2\theta\delta_{\mu\phi}\delta_{\nu\phi})
-g_{\mu\nu}\{\frac{\alpha^2(r+A)(r+B)}{2r^3(r+\alpha)^3}
-\frac{{2Q'_E}^2}{r(r+\alpha)^3}+\frac{{2Q'_M}^2}{r^3(r+\alpha)}
\} \label{4-37}
\end{equation}
Calculating the eigenvalues of $T_{\mu\nu}$ we can show that it
satisfies the dominant as well as the strong energy condition
outside and on the external horizon.

 The solution given by Eqs  (\ref{4-17})-(\ref{4-20})
has for $A<B<0$ and $\alpha>0$ the Hawking temperature

\[T_H=\frac{|{\lambda^2}'(-A)|}{4\pi} = \frac{B-A}{4\pi
A(A-\alpha)}\]
\begin{equation}
=\frac{\sqrt{(2M^2-{Q'_E}^2-{Q'_M}^2)^2 -4{Q'_E}^2{Q'_M}^2}}{4\pi
M\{ \sqrt{(2M^2-{Q'_E}^2-{Q'_M}^2)^2
-4{Q'_E}^2{Q'_M}^2}+2M^2-{Q'_E}^2 -{Q'_M}^2\}} \label{4-38}
\end{equation}

Also in the general dyonic solution we can put
\begin{equation}
c=0,\>\>\>\>\>\>g''_1=\frac{1}{2}\>\>\>\>\>\>\mbox{and}\>\>\>\>\>\>g_3=-4g''_1
g''_2 \label{4-39}
\end{equation}
where as before $g''_1$ and $g''_2$ denote the values of $g'_1$
and $g'_2$ if $c=0$. Then Eq (\ref{4-1}) gives
\begin{equation}
f_{+}(\psi)=e^{\psi} \label{4-40}
\end{equation}
and Eqs (\ref{3-5}), (\ref{3-8}) and (\ref{4-2})-(\ref{4-5})
become
\begin{equation}
e^{\psi}=e^{\psi_{0}}(1+\frac{\alpha}{r})
 \label{4-41}
\end{equation}

\begin{equation}
ds^{2}=-\frac{(r+A)(r+B)}{r(r+\alpha)}dt^{2}
+\frac{r(r+\alpha)}{(r+A)(r+B)}dr^{2} +r(r+\alpha)d{\Omega}
 \label{4-42}
\end{equation}

\begin{equation}
F_{+}=\frac{Q_Ee^{-\psi_0}}{(r+\alpha)^2}dr \wedge dt +Q_M
sin\theta d\theta \wedge d\phi
 \label{4-43}
\end{equation}

\begin{equation}
\frac{AB
e^{-\psi_0}}{2{Q_M}^2}=\frac{(\alpha-A)(\alpha-B)e^{\psi_0}}{2{Q_E}^2}=1
 \label{4-44}
\end{equation}

In the above solution let us make the replacements
\[r \rightarrow r'=r+\alpha,\>\>\> A\rightarrow A'=A-\alpha,\>\>\> B\rightarrow
B'=B-\alpha,\]
\begin{equation}
\alpha\rightarrow \alpha'=-\alpha,\>\>\psi\rightarrow
\psi'=-\psi,\>\>\psi_0\rightarrow \psi'_0=-\psi_0 \label{4-45}
\end{equation}
Then we find that the solution given by Eqs
(\ref{4-41})-(\ref{4-44}) in the primed quantities is identical to
the solution given by Eqs (\ref{4-17})- (\ref{4-20})

Finally in the general dyonic solution let as put
\begin{equation}
c=0\>\>\>\>\mbox{and}\>\>\>\>2g''_1 g''_2+g_3=0 \label{4-46}
\end{equation}
where $g''_1$ and $g''_2$ denote the values of $g'_1$ and $g'_2$
if $c=0$. Then Eq. (\ref{4-1}) gives
\begin{equation}
f_{\pm}(\psi)=g''_1 e^\psi+g''_2 e^{-\psi} \pm \sqrt{{g''_1}^2
e^{2\psi}+{g''_2}^2 e^{-2\psi}}
 \label{4-47}
\end{equation}
Eqs (\ref{3-5}), (\ref{3-8}), (\ref{4-3}) and (\ref{4-4}) become
Eqs (\ref{4-17}), (\ref{4-18}) and (\ref{4-10}), and Eqs
(\ref{4-2}) and (\ref{4-5}) take respectively the forms
\[F_{\pm}=\frac{Q_E}{\xi^2 f_\pm}dr \wedge dt+Q_Msin\theta d\theta \wedge
d\phi=\frac{Q_M^2}{Q_E}\{g''_1\frac{e^{\psi_{0}}}{r^2}\]
\begin{equation}
+g''_2\frac{e^{-\psi_{0}}}{(r+\alpha)^2}\mp
\sqrt{{g''_1}^2\frac{e^{2\psi_{0}}}{r^4}+
{g''_2}^2\frac{e^{-2\psi_{0}}}{(r+\alpha)^4} } \}dr \wedge dt+Q_M
sin\theta \wedge d\phi
 \label{4-48}
\end{equation}
\begin{equation}
{Q_E}^2=2g''_1 g''_2 {Q_M}^2
 \label{4-49}
\end{equation}
where $g''_1$ and $g''_2$ are given by Eqs (\ref{4-10}).If we use
Eq. (\ref{4-49}) to determine $Q_E$ we can say that we have the
parameters $A$, $B$, $\alpha$, $\psi_0$ and $Q_M$, which must
satisfy Eqs (\ref{4-10}). Therefore if we introduce the ADM mass
$M$ by Eq. (\ref{4-22}) we can take $M$, $Q_M$ and $\psi_0$ as
free parameters, and express $A$, $B$ and $\alpha$ by Eqs
(\ref{4-11}) - (\ref{4-13}). Then according to Ref. \cite{Nu} the
above is a "hairy" solution. The "hair" is "primary".

\end{document}